\documentclass[a4paper,11pt]{article}
\usepackage{pos}
\usepackage{xspace}
\usepackage{wrapfig}
\usepackage{enumitem}
\usepackage{lineno}
\usepackage[capitalize]{cleveref}
\usepackage{siunitx}

\title{Probing hadronic interactions at the 100 TeV scale with the Pierre Auger Observatory}
\ShortTitle{Probing hadronic interactions at the 100 TeV scale with the Pierre Auger Observatory}

\author*[a]{Eva dos Santos}
\onbehalf{for the Pierre Auger Collaboration$^b$}
\affiliation[a]{FZU - Institute of Physics of the Czech Academy of Sciences, Na Slovance 1999/2, Prague, Czech Republic}

\affiliation[b]{Observatorio Pierre Auger,
Av.\ San Mart{\'\i}n Norte 304, 5613 Malarg\"ue, Argentina \\
Full author list: \normalfont{\url{https://www.auger.org/archive/authors\_2025\_08.html}}}

\emailAdd{spokespersons@auger.org}

\newcommand{\Xmax}{X_{\mathrm{max}}} 
\newcommand{\Sthousand}{S \left(1000\right)}

\abstract{Extensive air showers produced by the interaction of ultra-high-energy cosmic rays 
($E > \SI[scientific-notation=true, retain-unity-mantissa=false]{1e18}{\electronvolt}$) in the Earth's
atmosphere provide a challenging yet unique channel to probe hadronic interactions at the 
$\SI{100}{\tera\electronvolt}$ center-of-mass energy scale.
Over more than 20 years of operation, the Pierre Auger Observatory has delivered invaluable insights into the 
modeling of hadronic interactions at energies beyond human-made particle accelerators. 
Notably, predictions from current models of hadronic interactions yield a muon deficit that becomes more 
pronounced with energy when compared to measurements. Presently, the interpretation of the nuclear mass 
composition estimated from the muon content is in tension with that from direct measurements of the depth of 
the maximum of electromagnetic profiles. 
Yet, the measured fluctuations of the muon content of air showers are in agreement with model predictions. 
These findings hint at small deviations in hadronic models that accumulate throughout the whole shower 
development rather than at large errors in the calculation of the first hadronic interactions. 
Also, in an independent, data-driven analysis, we show that the muon deficit can be alleviated if we allow a 
shift of the predicted depth of the maximum of air-shower profiles by 
$\qtyrange{30}{50}{\gram\per\centi\metre\squared}$ towards a heavier mass composition. 
More recently, we have also provided an updated measurement of the proton-proton cross-section at a center-of-mass
energy of $\SI{57}{\tera\electronvolt}$ and the first estimates of the neutron content of air showers by exploiting
late-time signals from the surface scintillator detectors of AugerPrime, the present upgrade of the 
Observatory. With the advent of AugerPrime, we expect to deliver breakthrough results on the 
\SI{100}{\tera\electronvolt}--scale hadronic interactions in the next decade.
}

\FullConference{19th International Conference on Topics in Astroparticle and Underground Physics (TAUP2025)\\
24–30 Aug 2025\\
Xichang, China\\}

\begin{document}
\maketitle

\section{Introduction}
\vspace{-0.4em}
For energies above $\SI[scientific-notation=true, retain-unity-mantissa=false]{1e15}{\electronvolt}$, the cosmic ray flux becomes 
too low to allow direct detection. 
In this regime, the properties of cosmic rays are inferred from comparisons between Monte Carlo simulations and data, using 
several hadronic interaction models that describe the development of extensive air showers in the atmosphere and other 
media. 
At $\SI[scientific-notation=true, retain-unity-mantissa=false]{1e17}{\electronvolt}$, the center-of-mass energy of a cosmic ray 
interacting in the atmosphere is equal to the current nominal energy of the Large Hadron Collider, becoming of the order of a 
few hundred TeV for $\SI[scientific-notation=true, retain-unity-mantissa=false]{1e19}{\electronvolt}$. 
In the absence of a theory that allows quantum chromodynamics to be calculated from first principles for very forward 
pseudorapidity regimes, where the most relevant interactions occur, and where there is a lack of experimental data to 
constrain the models, combined with an energy extrapolation of about one order of magnitude, hadronic interaction models 
become one of the most significant sources of systematic uncertainties in cosmic ray measurements at the highest 
energies~\cite{Reininghaus:2021zge}. 

\section{Pierre Auger Observatory}
\vspace{-0.4em}
The Pierre Auger Observatory, located near the town of Malarg\"ue, in the province of Mendoza, Argentina, is the world's 
largest cosmic-ray detector. 
It has a hybrid design comprising a $\SI{3000}{\kilo\metre\squared}$ surface detector (SD) array of 1660 water-Cherenkov 
detectors and the fluorescence detector (FD), which consists of a set of 27 fluorescence telescopes placed at four locations 
overlooking the SD array, operating on clear, moonless nights with a duty cycle of about 15\%~\cite{PierreAuger:2015eyc}. 
Events detected by at least one triggered SD station and at least one fluorescence telescope are called hybrid. 
From these, several analyses are based on a subsample of high-quality hybrid events that independently triggered the SD array 
and the FD, pass a set of strict reconstruction quality criteria, and are used for the energy 
cross-calibration of the SD array and have the highest reconstruction resolution.
At the Pierre Auger Observatory, we have conducted a set of independent analyses to measure the most relevant properties of 
hadronic interactions and have published model-independent methods that allow for testing current model predictions. 
With the advent of AugerPrime, the major upgrade of the Observatory, we strive to further reduce systematic uncertainties by 
using multi-hybrid events~\cite{PierreAuger:2016qzd}, which combine several detection techniques to enhance our estimation 
of the cosmic-ray mass composition.

Below, we summarize the main results from our analyses targeting several measurements of hadronic interactions at 
$\sqrt{s} \gtrsim \SI{100}{\tera\electronvolt}$.

\section{Exploring the 100 TeV energy scale with the Pierre Auger Observatory}
\vspace{-0.2em}
\paragraph{Inelastic proton-proton cross-section:} Recently, we updated our measurement of the proton-proton cross-section for 
\mbox{$\sqrt{s} \geq \SI{40}{\tera\electronvolt}$} using a novel method that simultaneously estimates the 
proton-proton interaction cross-section and the primary cosmic-ray mass composition with reduced experimental 
and theoretical systematic uncertainties~\cite{Tkachenko:2025coj}.
The used dataset comprises the entire Phase I hybrid events, using the same selection criteria as 
in~\cite{PierreAuger:2023kjt}. 
The methodology consists of a simultaneous best fit to the data, using templates of hadronic model interaction 
models with modified parameters.
\begin{wrapfigure}{r}{0.5\textwidth}
    \centering
    \vspace{-0.08cm}
    \includegraphics[width=.4\textwidth]{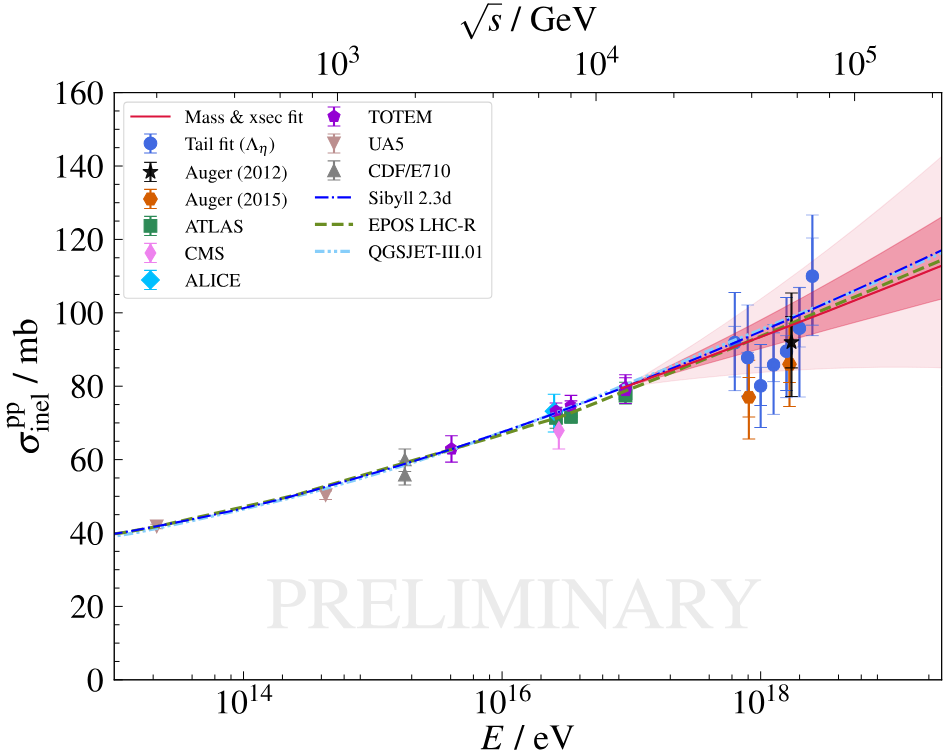}
    \caption{Inelastic proton-proton cross-section measurements for the current 
    (orange circles) and previous (black star) methods. Predictions from several hadronic model predictions and accelerator data are also shown. 
    The darker (light) shaded bands denote the statistical (total uncertainty)~\cite{Tkachenko:2025coj}.}
    \label{fig:1}
\end{wrapfigure}
The data comprise the full distribution of the depth of maximum development of air showers, $\Xmax$, and the models
were modified to accommodate both a systematic shift in $\Xmax$ and an energy-dependent factor 
$f_{19}^{pp}$ used to rescale the proton-proton cross-section at a reference energy of 
$\SI[scientific-notation=true, retain-unity-mantissa=false]{1e19}{\electronvolt}$. 
The overall best fit to the data for the full two-dimensional scan in $f_{19}^{pp}$, the proton-proton cross-section 
rescaling at the reference energy, and $\delta\Xmax$, the shift in $\Xmax$, was 
$f_{19}^{pp} = 0.97_{-0.07}^{+0.09} \mathrm{(stat.)}_{-0.18}^{+0.24}\mathrm{(sys.)}$, and a shift of 
$\delta\Xmax = -10_{-1.5}^{+1.7}\mathrm{(stat.)}_{-14.3}^{+13.6}\mathrm{(sys.)}\,\mathrm{g\,cm^{-2}}$. 
The obtained results for the inelastic proton-proton cross-section are shown in~\cref{fig:1} by the 
orange circles, and are compatible with the previous measurements (black star).
\vspace{-0.3cm}
\paragraph{Testing the electromagnetic and hadronic model predictions:} This method is data-driven and consists of a 
simultaneous two-dimensional fit to the distributions of $\Xmax$, the depth of the shower maximum, and $\Sthousand$, 
the signal measured by the water-Cherenkov detectors of the SD array at $\SI{1000}{\meter}$ from the shower axis. 
Due to the attenuation of the electromagnetic signal measured at the ground as a function of the zenith angle, the data are 
binned into several zenith-angle bins with similar numbers of events. 
The distributions of $\Xmax$ and $\Sthousand$ from data are compared with those from simulation templates using several 
hadronic interaction models and nuclear mass fractions contemplating four primary particle species. 
The best match between data and simulations is estimated via a maximum-likelihood fit, leaving the $\Xmax$ and $\Sthousand$
scales as free fit parameters~\cite{PierreAuger:2024neu}. 
Using the full data set of the highest-quality Phase I hybrid events with zenith angles below 
$60^{\circ}$, in the 
$10^{18.5}\,\si{\electronvolt}$ - $\SI[scientific-notation=true, retain-unity-mantissa=false]{1e19}{\electronvolt}$ 
energy range, the best description of the data is achieved if model predictions simultaneous allow for $\Xmax$ to be shifted 
towards deeper values, by about $\SI{20}{\gram\per\centi\metre\squared}$ ($\SI{50}{\gram\per\centi\metre\squared}$), 
and the muon content to be increased by $15\%$ ($25\%$) for EPOS-LHC (QGSJetII-04). 
These results imply that the estimated nuclear mass composition should be heavier than the one models predict. 
The new EPOS LHC-R predictions have a deeper $\Xmax$, already consistent with the data.  
However, EPOS LHC-R also shows a large variability in the hadronic signal with zenith angle, which should be 
corrected~\cite{PierreAuger:2025eiw}.
\vspace{-0.3cm}
\paragraph{Estimating the muon content at high-zenith angles:} Above $60^{\circ}$, most of the electromagnetic content of 
the shower is absorbed in the atmosphere, and the shower footprint at the ground becomes asymmetric and elongated due to 
the effect of the geomagnetic field. 
For these reasons, events with zenith angles above $62^{\circ}$ require a distinct reconstruction~\cite{PierreAuger:2014ucz}. 
The estimation of the average muon content $\langle R_{\mu} \rangle$ is derived from an overall scaling of the shower footprint 
signal relative to the predictions of 
$\SI[scientific-notation=true, retain-unity-mantissa=false]{1e19}{\electronvolt}$ proton-initiated showers using QGSJetII-03 
as the reference model.
\begin{figure}
    \centering
    \includegraphics[width=.3\textwidth]{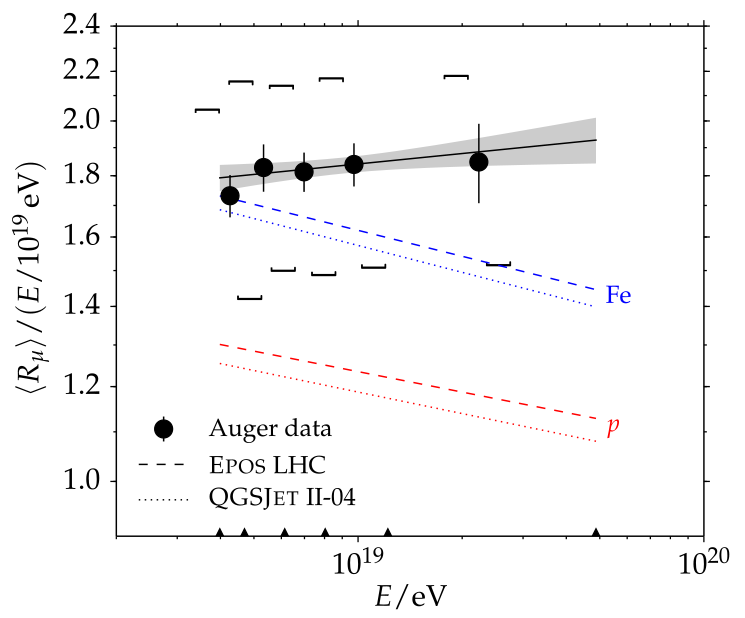}
    \includegraphics[width=.3\textwidth]{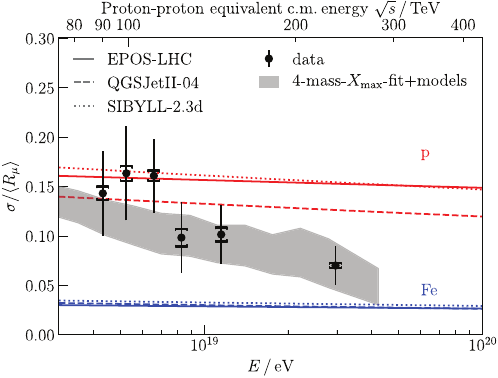}
    \includegraphics[width=.3\textwidth]{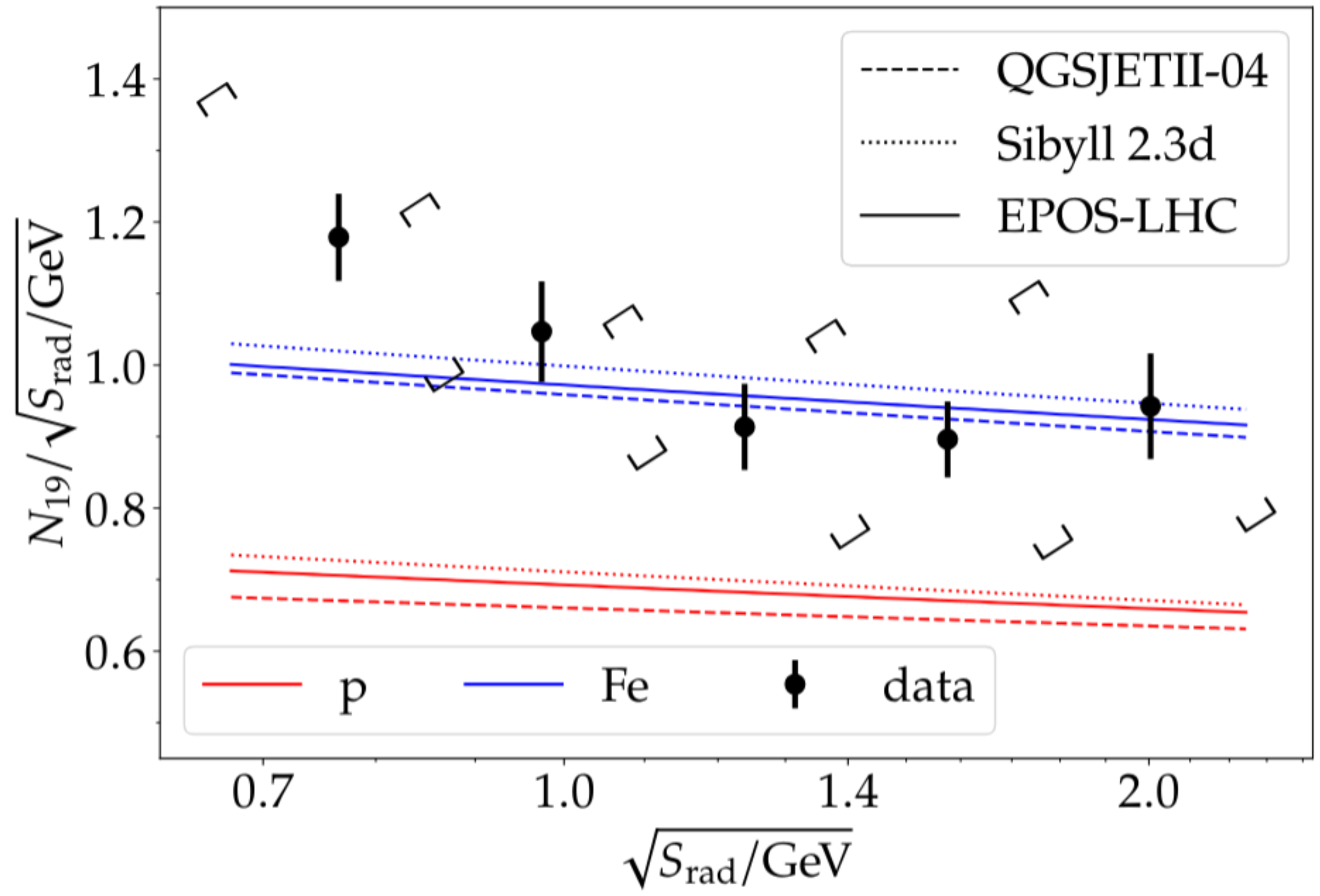}
    \caption{Average (left) and fluctuations (middle) of the muon content as a function of the shower energy for hybrid events, \cite{PierreAuger:2014ucz, PierreAuger:2021qsd}.
    Right: Same as the left panel, but replacing the FD by AERA,~\cite{PierreAuger:2025kym}.}
    \label{fig:2}
\end{figure}
To break the degeneracy between cosmic-ray energy and nuclear mass composition in the measured muon content, the energy estimate was obtained from the FD for our highest-quality hybrid events. 
The estimated mass composition obtained from $\langle R_{\mu} \rangle$~\cite{PierreAuger:2014ucz}, and the fluctuations 
$\sigma\left(\langle R_{\mu} \rangle \right)$~\cite{PierreAuger:2021qsd} are shown in the left and middle panels 
of~\cref{fig:2}. 
Finally, in the right panel of~\cref{fig:2}, using a new hybrid technique, the average muon content was estimated from 
40 high-quality events measured with the SD array and the Auger Engineering Radio Array 
(AERA)~\cite{PierreAuger:2012ker, PierreAuger:2025kym}. 
The small statistic results from the limited area of the AERA, of only $\SI{17}{\kilo\metre\squared}$. 
This work served as a successful proof-of-concept for the AugerPrime radio detector, demonstrating the ability to extend the 
duty cycle of hybrid events at high zenith angles. 
From inspection of the left and right panels of~\cref{fig:2}, we see that the measured muon content is above the model predictions 
for iron-initiated showers. 
However, the measured fluctuations in the muon number are bracketed by model predictions and are compatible with those from the 
electromagnetic longitudinal profiles.
These results suggest that the muon deficit may be due to small modifications in hadronic interactions that accumulate over 
many generations~\cite{PierreAuger:2021qsd}. 

\section{Summary and conclusions}
\vspace{-0.4em}
Ultra-high-energy cosmic rays are a challenging yet unique channel for probing hadronic interactions at
$\sqrt{s} \gtrsim \SI{100}{\tera\electronvolt}$. 
Auger results hint that hadronic interaction models still cannot provide a consistent description of extensive air showers, 
calling for improvements both in the hadronic and electromagnetic cascades.
With AugerPrime, we can further study the neutron component of air showers~\cite{PierreAuger:2025ibu} and reduce systematic 
uncertainties from hadronic interaction models using multi-hybrid events.

\paragraph{Acknowledgments:}
This work is funded by the Czech Ministry of Education, Youth and Sport, under the project CZ.02.01.01/00/22\_008/0004632.

\end{document}